\RequirePackage{etoolbox}
\csdef{input@path}%
{%
 {sty/},
 {img/},
}
\documentclass{elsevierbook}

\usepackage[sorting=none, maxnames=50]{biblatex}
\usepackage{algorithm}
\titleformat{\subsubsection} {\normalfont\fontfamily{phv}\fontsize{9}{10}\bfseries}{\thesubsection}{1em}{}

\begin{document}

\Frontmatter

\Mainmatter
  \begin{frontmatter}

\chapter{High Performance Computing and Computational Intelligence Applications with MultiChaos Perspective}\label{chap1}

\begin{aug}
\author[addressrefs={ad1,ad2}]%
{%
\fnm{Damiano} \snm{Perri}%
}%
\author[addressrefs={ad1,ad2}]%
{%
\fnm{Marco} \snm{Simonetti}%
}%
\author[addressrefs={ad2}]%
{%
\fnm{Osvaldo} \snm{Gervasi}%
}%
\author[addressrefs={ad2}]%
{%
\fnm{Sergio} \snm{Tasso}%
}
\address[id=ad1]%
{%
University of Florence,
Department of Mathematics and Computer Science ``Ulisse Dini'',
Viale Morgagni 67/a,
50134 Firenze (Italy)\\
}%
\address[id=ad2]%
{%
University of Perugia,
Department of Mathematics and Computer Science,
Via Vanvitelli 1,
06123 Perugia (Italy)
}

\end{aug}


\begin{abstract}
The experience of the COVID-19 pandemic, which has accelerated many chaotic processes in modern society, has highlighted in a very serious and urgent way the need to understand complex processes in order to achieve the common well-being. 
Modern High performance computing technologies, Quantum Computing, Computational Intelligence are shown to be extremely efficient and useful in safeguarding the fate of mankind. 
These technologies are the state of the art of IT evolution and are fundamental to be competitive and efficient today.
If a company is familiar with these techniques and technologies, will be able to deal with any unexpected and complicated scenario more efficiently and effectively.
The main contribution of our work is a set of best practices and case studies that can help the researcher address computationally complex problems.
We offer a range of software technologies, from high performance computing to machine learning and quantum computing, which represent today the state of the art to deal with extremely complex computational issues, driven by chaotic events and not easily predictable. 
In this chapter we analyse the different technologies and applications that will lead mankind to overcome this difficult moment, as well as to understand more and more deeply the profound aspects of very complex phenomena. 
In this environment of rising complexity, both in terms of technology, algorithms, and changing lifestyles, it is critical to emphasize the importance of achieving maximum efficiency and outcomes while protecting the integrity of everyone's personal data and respecting the human being as a whole.
\end{abstract}

\begin{keywords}
\kwd{High Performance Computing}
\kwd{Computational Intelligence}
\kwd{Machine Learning}
\kwd{Quantum Computing}
\kwd{Cloud Computing}
\kwd{Container}
\kwd{Neural Networks}
\kwd{Privacy}
\kwd{MultiChaos}

\end{keywords}

\end{frontmatter}

\section{Introduction}\label{Intro}
The pandemic caused by the COVID-19 virus has brought a strong wind of change to contemporary society. Our habits and the way we work have been transformed and we have had to adapt to a different lifestyle.
This abrupt transition has highlighted issues that require an evolution of IT services and infrastructures that adopt the state-of-the-art technologies that the research world offers. This is just the latest example in a long series of events that have forced companies to sharpen their computational techniques to cope with sudden, emerging needs.

Over the past year, we have all seen how videoconferencing has become widespread habit, smart working has become part of the daily routine, and the amount of data transiting the Internet has increased by about 20\% \cite{internetCovid}, and its use is not expected to decrease in the future.
This imposes a radical change in many areas, including that of computational resources, which will be increasingly important to tackle complex problems generated by multi-chaotic situations.

Many companies and organisations still use non-scalable architectures configured with a single general purpose server on which they perform calculations. 
This type of solution is not well suited to today's needs, where it is important to be able to meet unexpected peaks in load while ensuring high reliability, i.e. constant availability of the services offered, and redundancy in the event of failure.
The open source world provides a great deal of extremely interesting software that, for example, makes it possible to create clusters of computer nodes that adapt the computational capacity provided through various techniques, first and foremost horizontal scaling.

The paper is articulated as follows: in section \ref{relwork} the most relevant works, in our opinion, that have dealt with the aspects we have outlined in this paper are presented. In section \ref{hpc} we describe the approaches that we consider most significant for solving complex problems that require a high number of operations and which can therefore benefit from advanced techniques for dividing the work into parallel computing nodes running cloud containers, exploiting the GPGPU words and neural networks.
In section \ref{quantum} we describe how quantum computing today represents a cutting-edge frontier that will become increasingly accessible in the medium term, whose technical potential can be exploited to provide and improve existing software and services to citizens. 
In the coming years, this technology will prove to be the queen of technologies for reducing the complexity of problems arising in multi-chaos situations.
Section \ref{computationalIntelligence} describes some of the techniques that make it possible to accelerate the resolution of complex problems by exploiting the enormous potential offered by Machine learnings techniques (e.g. Multilayer Perceptron and Convolutional Neural Network) which, thanks to modern graphics accelerators, can reach dimensions that were unimaginable just a few years ago.
Section \ref{multichaos} addresses the increasingly pressing issue of personal data processing and its implications in chaotic situations, with reference to European legislation that, in addition to globally regulating the issue through the GPDR, is attempting to address the same issue in the Artificial Intelligence field, including what is presented in the Machine Learning and Neural Networks fields.

\section{Related Works}\label{relwork}
The abrupt and unstoppable computerization over the last 40 years has enabled the handling and management of massive amounts of data in ever-shorter time frames. Furthermore, the complexity of scientific and technical issues \cite{golub2014scientific}, as well as the automation of industrial and business processes, demanded an extremely sophisticated use of computational resources \cite{padua2011encyclopedia}. 
This required a variety of solutions, ranging from the development of CPUs, GPUs, TPUs, and other more powerful hardware components to the building of computer clusters capable of working on the same issue in parallel \cite{rashid2018distributed, takizawa2006hierarchical, kindratenko2009gpu}.

The study of algorithms that are less greedy for computational resources can significantly contribute to making information structures more efficient; in many application problems in the field of engineering and logistics \cite{chassiakos2019evolutionary, ghorbani2018optimizing, yan2017hybrid, abbasi2020efficient}, classical analytical methods for the efficient search of the maximum or minimum values of an objective function have been supported by very effective heuristic and meta-heuristic solutions.

Other times, the problems have an inherent complexity that makes them intractable even with extraordinary resources, such as factorization of extremely large integers or finding solutions for a certain kind of PDEs (Partial Differential Equations). A paradigm shift is required in such cases \cite{caleffi2018quantum, bermejo2018architectures, childs2018toward}.

Artificial intelligence technologies are crucial in environments with high computational complexity, such as chemistry and mathematical-scientific modeling\cite{DBLP:conf/iccsa/LaganaGTPF18,DBLP:conf/iccsa/PerriSLLG20}; it is also import in the recognition of specific patterns in images or other data aggregates, such as tumor recognition or emotion recognition directly from images or photos\cite{DBLP:conf/iccsa/BenedettiPSGRF20,DBLP:conf/iccsa/BiondiFGP19,DBLP:conf/iccsa/FranzoniTPP19}.

The management of scenarios in which the system variables are numerous and change rapidly assumes an automatic, and in some cases intelligent, adaptation by the calculation structure \cite{ho1992short, li2009improved, georgakopoulos2017novel}.
Recently, the rapid rise of Artificial Intelligence techniques has enabled us to increase the responsiveness of decision-making processes, as well as to introduce forms of automation and relational empathy with machines at previously unthinkable levels \cite{chakrabarty2020secure, gao2019generating}.

There is no doubt that such extensive technical progress has a significant influence on society: smart cities, which open their doors to extraordinary views and scenarios for the possibilities of services given to their citizens, to well-founded concerns about their environmental impact \cite{bibri2017social}; smart schools and educational institutions, that have made significant investments in technology equipment to improve teaching efficacy, whose true outcomes and advantages are still a hotly debated topic among many researchers \cite{surry2010technology, BULMAN2016239, chauhan2017meta,DBLP:conf/iccsa/SimonettiPAG20,AUYONGOLIVEIRA2018954, perri2021learning}; smart working is a new form of working that technology has made easier and leaner, and which has proven to be appropriate for some working situations, but whose true usefulness in circumstances with a high degree of complexity and unpredictability has yet to be established \cite{mcewan2016smart, angelici2020smart}.

As a result, we can state that the extraordinary complexity of today's society drives us to seek new solutions, whose significant implications may not be readily apparent. 
It is recommended that new models be developed to assist us in the interpretation of signals, comprehension of the implications of very quick changes, and management of both normal and emergency situations. The construction of such models necessitates a significant amount of study, which must be backed up by powerful and rapid technological support.

\section{High Performance Computing approaches to solving complex problems}\label{hpc}

The pandemic is only the most recent and scorching example of a multi-chaotic situation with a devastating impact at planetary level, which in a very short time, thanks to the extraordinary research carried out before and during the period under consideration, by various companies and research groups, has managed to produce different types of vaccines that will hopefully bring us out of the stagnant waters in which we have had to stay for an extended period of time.

In this section, we present a number of technologies that we believe are fundamental resources for modern and near future research and will enable us to reduce the complexity inherent in various future scenarios. 

\subsection{Cloud containers}

Cloud Computing technologies have advanced dramatically in recent years, progressing from the supervision of physical computers to the virtualisation of environments and, eventually, the use of containers, resulting in improved application administration and environment separation.

Containers facilitate cloud infrastructures, and even personal computer resources, to be more flexible by enabling horizontal scalability and, as a result, allowing the infrastructure to be tuned to the application demand. 
These considerations are critical in order to best calibrate the resources required to address the complexities of the present challenges.

The tested collection of products and technologies may be customised and adapted to the demands of any organisation or corporation that wants to manage data and infrastructure by establishing a private or hybrid Cloud environment (i.e. able to use computing resources from a provider of commercial Cloud services).\cite{iccsa/cloud2021impress}

We will show how open source components may be used to build a highly dependable distributed system with security, quick horizontal scalability, and the capacity to be easily extended to commercial Clouds.

For demonstration purposes, we will discuss the key procedures for establishing a cluster comprised of nodes located throughout the company's offices. The technology employed will be container-based, thanks to the utilisation of Docker Engine. 
The different containers are orchestrated by running Docker in a specific mode for natively controlling a cluster of Docker Engines known as a swarm. 
It is possible to build a swarm, deploy application services to a swarm, and govern swarm behaviour using the Docker Command Line Interface. 
The cluster will be made up of four nodes, with a minimum Linux distribution deployed on each.

\subsection{Containers insights}\label{containers}

Data volatility is a major concern when deploying replicated containers.
Containers, in reality, are stateless, and the data contained within them is lost the instant the container is destroyed.
The destruction of a container might also occur in an unforeseen manner; for example, it is conceivable that autoscaling identifies low system usage and decreases the number of active nodes as a result.
To tackle this challenge, it is critical to provide a network-distributed file system that is subsequently used as storage by containers.
This ensures the permanence and consistency of information.
XFS\footnote{XFS (Extents File System) is a 64-bit, high performance journaling file system for Linux. It was initially created by Silicon Graphics for its IRIX OS, but the code was later donated to Linux. XFS works extremely well with large files and it is known for its robustness and speed. XFS supports filenames of up to 255 bytes, files of up to 8 EB and file systems of up to 16 EB.} is one of the file systems that may be utilised, and its maintenance may be delegated to the GlusterFS program.
GlusterFS also allows you to provide a list of trustworthy servers.
These will create the trustworthy pool, which will be used to share available disk space across nodes.
Persistent data will uniformly be mounted on each node.
An enterprise grade fiber connection between sites with a speed of at least 1 Gbps will be required.

If our services need the usage of a database, this will be the component to be optimised next.
The most successful method entails the establishment of several containers to which certain tasks are assigned.
One of them will be the DB container, with the role of master, while the others will be slaves.
The master is used for writing and is the only one that can change the structure and data in the database.
Instead, slaves will be containers with only the necessary rights to perform reading queries.

The significant advantage is that we can add as many slave nodes as we need, improving our ability to satisfy customer demands linearly.
In general, individuals who explore a website perform far more reading actions than writing actions.

With this perspective, you can also create a database autoscaling mechanism that raises the number of reader nodes based on various factors such as average database usage percentage or average number of active requests per second.
There are two critical points to consider.

To begin with, there is some delay across the slave nodes (readers) and the master (writer) in this sort of design.
This means that there may be instances where reads on newly written data provide old and out-of-date results.
When a database is properly configured, latency is very low, often less than 100ms and mostly closer to 30ms or 40ms.

The second thing to keep in mind is that queries are not automatically sorted across master and slave nodes.

There are two methods to distribute the workload among the writer and the readers.
The first is to act on the program code that utilises the database, which requires creating two access points to the database and sorting queries that only perform reading operations on the access point to the slaves and queries that only execute writing operations on the access point to the master.
This solution, however, is not always feasible, in fact it is perfectly plausible that the program used is legacy or closed source.

The second method, which arises as an alternative to this problem, requires to define an additional container that acts as a load balancer and automatically sorts requests across nodes.
In the case of a MariaDB database for instance, Maxscale proxy can be used.
In this regard, we are going to publish the port of MariaDB MaxScale REST API, an HTTP interface, which generates data in JSON format, offering visual management tools. 
The access to the proxy implies the addition of a new rule to the NAT Network port forwarding table.
MariaDB MaxScale divides requests such that writing queries are sent to the master container, while reading queries are handled and balanced through the $slave1, slave2, \dots  , slaveN$ containers.
It is also feasible to include the master in the pool of nodes capable of reading.
Then, in the master, a special user with \texttt{GRANT ALL} access must be defined for MaxScale.
MariaDB MaxScale is released under a BSL (Business Source License) license and is capable of much more than simply load balancing: it can also handle failover and switchover.
A few stages are involved in the configuration: establishing the servers, creating the monitoring service, defining traffic routing using the Read-Write-Split mode, and lastly configuring the listener agents using the \textit{mariadbclient} protocol and its TCP port.

The next container we are going to analyse is the proxy for web services.
It exposes port 80 for the HTTP protocol and port 443 for the HTTPS protocol.
Its task will be to receive and sort requests coming from the network and redirect them to right web servers (such as Apache, NodeJS or Nginx).
\begin{figure}[ht]
    \centering
    \includegraphics[width=\linewidth]{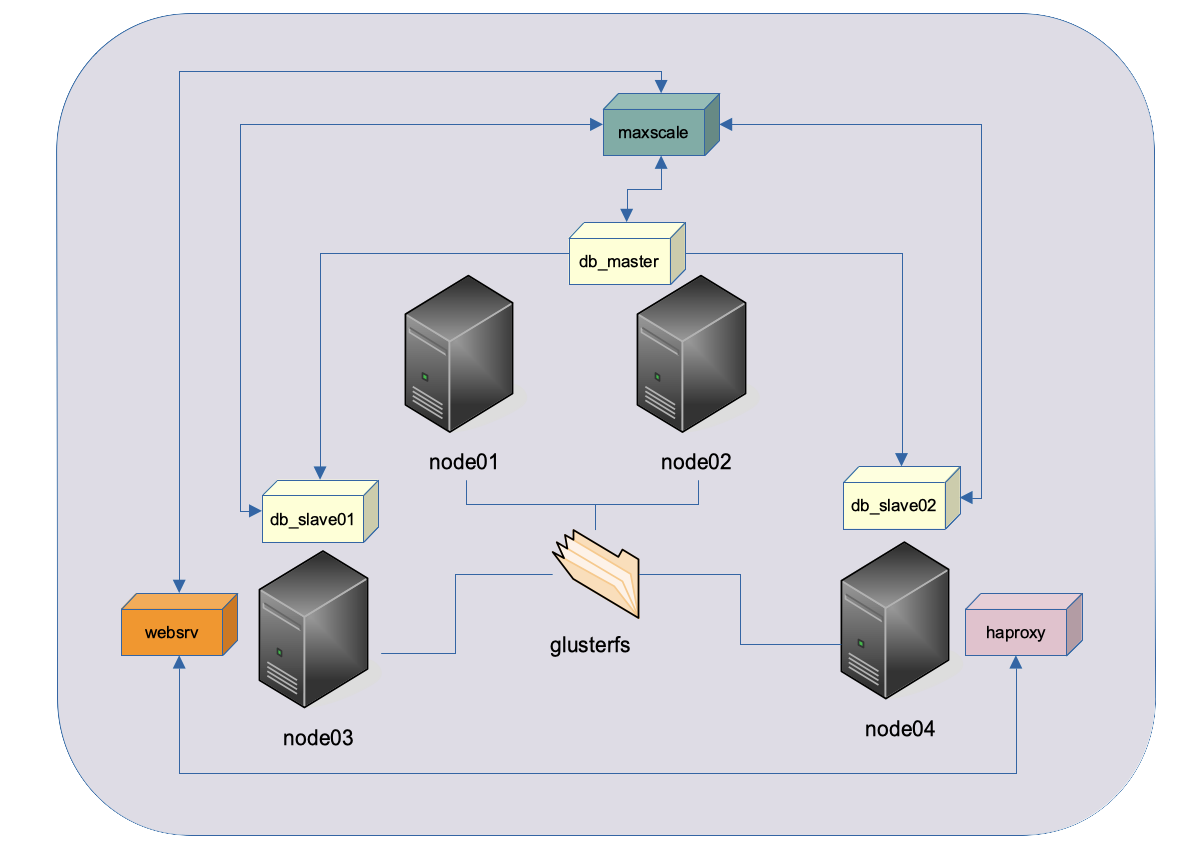}
    \caption{Final Architecture}
    \label{fig:containers}
\end{figure}
To appropriately configure the proxy, the documentation given by the HAProxy official site is consulted - The Reliable, High Performance TCP/HTTP Load Balancer\footnote{See the URL: \url{https://hub.docker.com/_/haproxy}}, where it is explained the use of the software in a Docker Swarm environment.
The DNS resolver definitions are very critical elements to consider.
These are required for identifying the containers that comprise the back-end.
Furthermore, proper request timeout settings must be defined and calculated to reduce the chance that some blocked processes will bind connections, causing difficulties that would render services inaccessible to users.
The HTTP mode setting enables HAproxy to make sophisticated decisions, such as routing traffic within a certain group of servers depending on the URL route requested by clients, as well as routing traffic depending on the HTTP headers received.
In fact, this parameter allows HAproxy to work as a level 7 load balancer.
This mode is essential for load balancing.

It is also necessary to specify the relevant certificates for the use of the HTTPS protocol or alternatively we can rely on other containers that deal with the automatic generation of certificates, such as those developed by EFF Let's Encrypt\cite{LetsEncrypt}.
In the back-end section we will establish the algorithm for balancing requests to Web servers.
There are several algorithms, the most common is \textit{Round Robin}, which will select the recipient servers one by one in a cyclic way.

Alternatively, the \textit{LeastConn} algorithm may be used: it chooses the Web server that will handle the request, depending on the number of connections remaining active and always selects the Web server with the minimum load.
With this final container, the entire infrastructure may be completed.

The completed infrastructure is depicted in Figure \ref{fig:containers}.
We can check the status and distribution of all containers built using the container orchestration tool in swarm mode, Docker service, which is only executable by the management node.
Finally, it is vital to note that Portainer\footnote{Portainer is a universal container management tool. It works with Kubernetes, Docker, Docker Swarm and Azure ACI. It enables to manage containers without needing to know platform-specific code. See the URL \url{https://www.portainer.io/}.} may be used to manage and administrate all of the containers that have been built.

It has a web interface that makes long-term system maintenance easier, as well as access to logs, status graphs, and tools for restarting and restoring individual containers.

\subsection{GPGPU Computing}\label{gpgpucomputing}
GPGPU is an abbreviation that stands for general-purpose computing on graphics processing units. 
This term refers to the usage of graphics accelerators, which are essential components of computer graphics, for generalised mathematical operations and calculations that result in the acceleration of computations. 
With the advent of customizable shaders and support for floating point computations in 2001, GPGPU computing grew in popularity.
The capacity to carry easily out SIMD (single instruction on multiple data) calculations, is one of the features of these architectures.

GPGPU computing became easily useful even in the consumer environment in 2006, owing to the launch by NVIDIA of the 8800 series of graphics cards with G80 processor, which could benefit from CUDA\footnote{CUDA is a proprietary framework released by NVIDIA\texttrademark}, an architecture capable of executing highly efficient programs for parallel computing on GPU.
In 2006, an 8800GTX with the G80 featured 128 unified shaders (compute units) working at a frequency of 576 MHz, which guaranteed 345.6 GFLOPS.
Today, 15 years later, the top-of-the-line card named RTX 3090 for desktop PCs created by nVidia delivers 8704 compute units at 1440Mhz that promise 35581 GFLOPS: 102.95 times larger than what was supplied by the finest video card in 2006.
 
\begin{figure}[!ht]
    \centering
    \includegraphics[width=\linewidth]{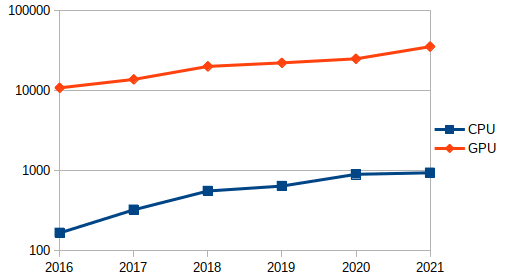}
    \caption{CPU vs GPU, performance in GFLOPS}
    \label{fig:gpucpu}
\end{figure}
The graph in Figure \ref{fig:gpucpu} depicts the rise in computing capability of CPUs and GPUs over time.
On the Y axis, the number of billions of floating point operations per second (GFLOPS) is reported.
The performance disparity is increasing year after year, as seen by the logarithmic scale.
GPUs become programmable architectures, similar to CPUs, thanks to both proprietary (CUDA) and Open Frameworks (OpenCL\footnote{OpenCL™ (Open Computing Language) is an open, royalty-free standard for cross-platform, heterogeneous, parallel programming devices (CPUs, GPUs, FPGAs, DPSs} and 
SYCL\footnote{SYCL is a royalty-free, cross-platform abstraction layer that enables code for heterogeneous processors to be written using standard ISO C++ with the host and kernel code for an application contained in the same source file.
}), and can operate hundreds of threads at the same time.
More properly, since we are considering Compute Units, we can name this approach Single Instruction Multiple Thread (name given by NVIDIA to this architecture, SIMT).

CUDA and both OpenCL and SYCL have a strong relationship. 
The latter is the Open Source alternative to CUDA and enable to build parallel, in particular for a diverse collection of GPUs, including Intel, NVIDIA, and AMD.
However, the performance is not identical, and OpenCL, because to its feature that allows it to operate on many types of hardware, is often slower than CUDA.
In 2011, the performance gap was around 16\% \cite{cudavsopencl,cudavsopencl2} and to date it appears to have gotten even more pronounced \cite{cudavsopencl3}.

\subsection{GPGPU insights}
As mentioned in previous sections, the use of GPGPU computing can greatly speed up many algorithms.
Examples include cryptography, image and sound manipulation and analysis, multimedia coding and decoding, neural networks, and Natural Language Processing applications that are crucial in many areas today.
However, the use of these techniques, which represent the state of the art of research, requires the adoption of certain guidelines to lead to the expected results.

The architecture of the GPGPU involves running a host program on the CPU and a kernel program on the innumerable Compute Units of the GPU.
The host program has the function of allocating the necessary resources, monitoring the execution status of the kernels and collecting the final results.
Each Kernel will execute the required algorithm on a Compute Unit, acting on a portion of the input data.
The input data must be moved from the central memory to the GPU memory before Kernels are executed and at the end of the kernel computation the output results has to be copied from the GPU memory to the central memory.
This data movement must be optimised, in the sense that has to be minimised the input-output requests outside the GPU memory. Furthermore, it is of fundamental importance to allocate data in such a way as to make maximum use of the memory areas close to the Compute Units, in line with the Memory model implemented by OpenCl.
The data transfer has a computational cost both in terms of time and machine cycles and can have a heavy impact on the measured performance.
We can conclude that the use of these technologies is optimal in the case of medium to large input data.\cite{AES2010}

Although the use of these graphics accelerators results in a considerable increase in power, the performance-per-watt ratio\footnote{It evaluates the amount of calculation a computer can perform for every watt of power consumed.} is in favour of GPUs over CPUs.\cite{gpuPerformancePerWatt}
For example, the Intel i9-11900K processor that hit the market in Q1 2021 has a Thermal Design Power (TDP \footnote{The maximum amount of heat created by a computer chip or component is known as the thermal design power (TDP), which is also known as the thermal design point.})  of 125 Watts and is capable of delivering 996.0 GFLOPS in SGEMM calculations.
The NVIDIA 3090 has a TDP of 350 Watts and is capable of delivering 35581 GFLOPS.
If we calculate the performance-per-watt ratio, we get 996/125=7.968 GFLOPS/watt for the CPU and 35581/350=101.66 GFLOPS/watt for the GPU.

\subsection{GPGPU and Neural Networks}

The field of machine learning has witnessed a boom of interest in recent years, since a number of studies have demonstrated the efficacy of neural networks in a variety of tasks that were previously thought to be extremely difficult. 
Experts have access to massive volumes of data, and contemporary accelerators have enormous computing capability.\cite{10.1145/3318170.3318183}

Developing code that solves a problem and makes it useful on many graphics card architectures, on the other hand, has become incredibly challenging, especially from an algorithmic standpoint.
This is owing to the fact that while programming on GPU, it is required to declare how memory and registers are to be used and allocated in great detail.
It should be noted that even graphics cards of the same brand and generation (but different model) might have significant variances.
Auto-tuning approaches have somewhat reduced the difficulty of performance portability by customizing memory structures and loops to a specific architecture.
Today there are some auto-tuning software and libraries that have been developed to cater to researchers powerful instruments and attempt to mitigate this problem.
There are various types of approaches that can allow to optimize kernel parameters for various GPU architectures, i.e. using a deterministic approach and the "Generate and Test" technique.
In more recent times, through the use of machine learning, neural networks have been created that are able to predict the best parameters to be used in a GPU code and optimize final performance.

GPGPU computing is currently used to accelerate many of the most common deep learning frameworks, such as TensorFlow, PyTorch, Caffe, Matlab, and others.
It is also used in linear algebra, Data analysis, cryptography, affective computing and so on\cite{DBLP:conf/iccsa/SantucciFACPS20}.
There are specific neural networks that work with images, and convolution is one of the most important operations.
Assume we have a picture saved on our device.
In most cases, a picture is represented using three colour channels: red, green, and blue (RGB), which are represented in memory with three distinct matrices with the same dimensions.
The number of rows will be equal to the height of the image, while the number of columns will be equal to the width of the image. 
Each matrix element will contain the integer that quantify the amount of Red, Green, Blu respectively.
Because most images use 8 bits to encode the brightness value (grey tone) of each pixel, these values are usually between 0 and 255. 
This results in 16 million colours being represented, as results combining 255 possible red tones, 255 possible blue tones, and 255 possible green tones.

Convolutional Neural Networks are a type of Deep Neural Network that performs well on data organised in a grid topology, such as time series and visual inputs, and are one of the most popular machine learning architectures for image-based classification. 
Image recognition, segmentation, detection, and retrieval are examples of applications where CNNs have achieved state-of-the-art results.
Convolutional effects can be varied, for example, enhancing edges, increasing contrast, dilation or eroding the area occupied by the objects in the image, and so on.
Winograd, Image to Column + GEMM, and Direct Convolution are the three main algorithmic techniques for performing convolution with digital images.
These three algorithms obtain the same output, but have different execution times.
Arrays of different sizes and parameters may be faster with the Winograd convolution than with the Direct convolution, and vice versa.
We demonstrated that it is possible to identify the best algorithm to adopt for the convolutional operations on the basis of the previous described arrays.\cite{cnnDamianoDividiti2019}.

\section{Quantum Computing to treat multichaos scenarios}\label{quantum}

The idea of using quantum mechanics in the world of computation was born around the 1950s following the statement by R. Feynman about the possibility of simulating nature efficiently and effectively.
Some years later, he described\cite{feynman2018simulating} the possibility of defining physical laws through a computer, by analyzing their probabilistic aspect: the important conclusion was that classical computation could not effectively simulate physical processes, that would only be reachable through a quantum computer.

In the late 1970s classical probabilistic computation became extremely important in computer science and as a result the first non-deterministic algorithms were implemented, giving rise to doubts about Church-Turing's thesis\cite{church1985articolo}.
This was followed by the studies of David Deutsch who in his\textit {"Quantum theory, the Church–Turing principle and the universal quantum computer"}\cite{deutsch1985quantum} lays the foundations to define how to apply quantum principles to the Turing machine.
He first talked about a quantum Turing Machine, that is, an abstract model that allowed to simulate a quantum computer.

In 1982, physicist Paul Benioff was able to demonstrate that the classical Turing machine could simulate certain physical phenomena without incurring an exponential slowdown in its performance. Three years later David Deutsch hypothesized that, since the laws of physics were ultimately approximated to those of quantum mechanics, a device based on the principles of quantum mechanics could be used to efficiently simulate an arbitrary physical system.
The true potential of this new science was highlighted in the early 1990s by Richard Jozsa, who in 1991, after describing the functions that cannot be solved by quantum parallelism, collaborated with Deutsch proposing the first problem that a quantum machine could solve more quickly than a deterministic one.

Later, in 1994, a mathematician named Peter Shor developed an algorithm that could find the factors of large numbers much more efficiently than the best classical algorithm. 
It was so powerful that it put modern cryptography at risk: in fact, encryption algorithms that until then (and even in the present day) were considered state-of-the-art, with the advent of this new type of computation would have soon become obsolete. It is thought that, given an integer N, Shor's algorithm can factorize it in a time of $O(log(N))$, while on a classical computer the time is exponential in N. This means that the quantum algorithm could easily pierce the best modern encryption algorithm, in a very short time.

Another notable algorithm was thought by the researcher Lov Grover, who was able to solve a search problem in a database of N unsorted items in $O(\sqrt{N})$ using $O(log N)$ as storage space. This is a real incredible result compared to classical search algorithm that operates in $O(N)$.

These advances contributed to the current definition of quantum computers, which are computers that use quantum physics and mechanics to provide computational power much superior to that of a classical computer for some types of problems.

The study of these machines has given rise to a new field of theoretical research in computer science and physics called \textit{quantum computation}, which will completely overturn the processing of information, managing to solve currently unsolvable scientific problems.
Despite the fact that the premises are very promising, the actual technical implementation for quantum computers has not helped us to achieve the desired results; in fact, managing a large number of qubits, which is required to solve problems of varying nature and structure, has demonstrated a significant level of implementation complexity. In addition, the realization of a quantum algorithm necessitates an ad-hoc configuration of the algorithm itself, due to the peculiarity of quantum transformations compared to classical ones, and the implementation of the corresponding quantum circuit.
Nonetheless, the time and resources required to design and construct such systems will be amply repaid with the ability to solve computationally very complex problems, which are currently difficult, if not impossible, to solve with existing hardware architectures.

\subsection{Bits and Qubits}\label{qubit}
\begin{figure}[!ht]
    \centering
    \includegraphics[width=\linewidth]{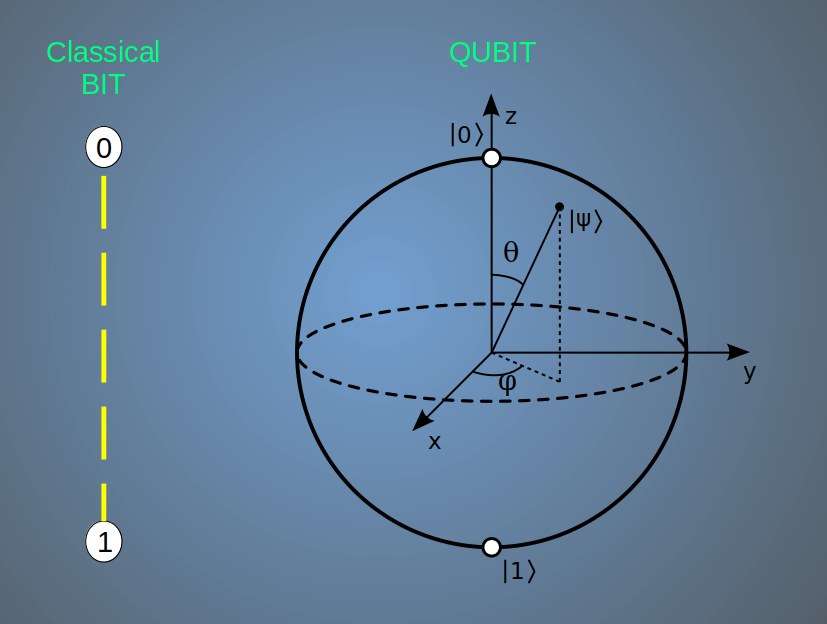}
    \caption{Bit and qubit}
    \label{fig:bit_qubit}
\end{figure}

The bit is the smallest unit of information; it is also the foundational element of Claude Shannon's classical theory of information, and it is physically equivalent to a two-state system with only two possible values: 0 and 1.
Instead, the quantum computer employs qubits (quantum bits), a physical property of an elementary particle that obeys quantum physics by simultaneously existing in two states: it can be both 1 and 0 at the same time, or even in a superposition of them, that is a linear combination of them.
A classical system could be compared to a region in real three-dimensional space made up of two distinct points, whereas a qubit could be compared to one of the infinite points on the surface of an unitary sphere (Figure \ref{fig:bit_qubit}).
This quantum physical device was defined by Benjamin Schumacher\cite{schumacher1995quantum} as being simpler, more flexible, and powerful than the digital one.
To comprehend the essence of this new kind of machine and how it differs from its traditional counterpart, the bit, we must consider the rules that govern the actions and development of a particular physical structure, and therefore the transmission of the knowledge found within it.
It has become possible to extend the field of operation of a machine by looking at it from a different point of view, thanks to some quantum mechanics postulates.
As a result, the space filled by classic binary sequences (bit registers) will contain all infinite variations (\textit{principle of superposition of states}) as well as their non-classical relations (\textit{phenomenon of interference}), and all of these will affect the final outcome (\textit{principle of measurement}).

Always referring to quantum physics we can identify the qubit as a vector $\psi$ in the two-dimensional complex space $\mathbb{C}^2$ (Hilbert space) defined in the form
\begin{equation}
    |\psi\rangle = \alpha|0\rangle + \beta|1\rangle =
    \alpha
    \begin{pmatrix}
        0\\
        1
    \end{pmatrix} +
    \beta
    \begin{pmatrix}
        1\\
        0
    \end{pmatrix},
\end{equation}
where the scalars $\alpha$ and $\beta$ are complex numbers expressing the probability amplitude of the state $|0\rangle$ and $|1\rangle$, respectively; $|0\rangle$ and $|1\rangle$ states constitute the computational basis for the space under consideration.
The probability that a qubit can be 0 or 1 is totally without guarantees and is dependent only on situations of uncertainty. These intermediate states of the qubit are considered \textit{superpositions of the states} and can be viewed as a coexistence of them in certain proportions: the probability that a qubit can be 0 or 1 is absolutely without guarantees and is based only on situations of uncertainty.

\textit{Entanglement} is another significant concept in considering quantum processes.
Entanglement is a state of correlation between two or more quantum systems in which two systems are related in a cause-and-effect relationship while not being in direct or indirect communication.
Following that, we may add \textit{quantum teleportation}, which is the instantaneous transmitting of a quantum state from one stage to another that is independent of the gap between the particles, which may be quite large.

\subsection{Quantum Register}\label{quantum_register}
It is possible to generate four possible states with two classic bits: 00, 01, 10, 11.
In general, $2^n$ distinct states can be constructed with n bits.
Each normalized vector in the state space created by a system of n qubits has dimension $2^n$ and represents a potential computational state, which we will refer to as the n qubit quantum register.

This exponential increase in the size of state space means that a quantum computer could process data at a rate that is exponentially faster than a classical computer.

A quantum register with \textit{n} qubits is formally a $2^{n}$ dimensional Hilbert space element, $\mathbb{C}^{2^n}$, with a computational basis generated by $2^{n}$ registers with \textit{n} qubits.
\begin{displaymath}
|{i_1}\rangle\otimes|{i_2}\rangle\otimes....\otimes|{i_n}\rangle \qquad i_j\in \{0,1\} \qquad 1\leq j\leq n.
\end{displaymath}
$|{i_1}\rangle|{i_2}\rangle....|{i_n}\rangle$, or more simply $|{i_1 i_2....i_n}\rangle$, is the name for a base vector (their set is known as \textit{computational basis}). As a result, using two qubits and the vectors $|{00}\rangle, |{01}\rangle, |{10}\rangle, |{11}\rangle$, we can create the computational foundation of the state space. The vector $|{01}\rangle$ can be written as $|{0}\rangle\otimes|{1}\rangle$, that is the tensor product of $|{0}\rangle$ e $|{1}\rangle$, as we've shown before.
\begin{displaymath}
|{0}\rangle = \begin{pmatrix}1\\0\end{pmatrix}   
\hspace{14pt}
|{1}\rangle = \begin{pmatrix}0\\1\end{pmatrix}
\qquad
|{01}\rangle = |{0}\rangle\otimes|{1}\rangle = \begin{pmatrix}0\\1\\0\\0\end{pmatrix}
\end{displaymath}
For instance, a two-qubit quantum register is characterized by a superposition of states, as seen in the equation below:
\begin{displaymath}
|{\psi}\rangle = \alpha_0|{00}\rangle+\alpha_1|{01}\rangle+\alpha_2|{10}\rangle+\alpha_3|{11}\rangle
\hspace{20pt}with\hspace{10pt}
\sum_{j=0}^3 |\alpha_j|^2=1
\end{displaymath}
It is precisely with the superposition of states that quantum parallelism occurs in all its power. Below we will analyze three quantum algorithms able to take an exponential advantage from the superposition of states:

\begin{itemize}[labelindent=1cm]
    \item the Deutsch–Jozsa algorithm
    \item the Grover's algorithm
    \item the Shor's algorithm
\end{itemize}

\subsection{Relevant quantum algorithms}
This section contains some important algorithms for quantum computing that will have a revolutionary impact on the world of computer applications in the next few years.

\subsubsection*{The Deutsch–Jozsa algorithm}\label{Deutsch_Jozsa}
In the Deutsch–Jozsa problem\cite{deutsch1992rapid}, the goal is to establish if an input unknown function has certain characteristics; this function, $f:\left\{0, 1\right\}^n\rightarrow \left\{0, 1\right\}$, with $n\in \mathbb{N}$, which works as a black block is called \textit{oracle}. This function can be either \textit{constant}, when all inputs are 0 or 1, or \textit{balanced}, if it has the value 1 for half of the input domain and 0 for the other half. The goal is to use the oracle in order to see whether $f$ is constant or balanced.\newline
\begin{figure}[!ht]
    \centering
    \includegraphics[width=0.7\linewidth]{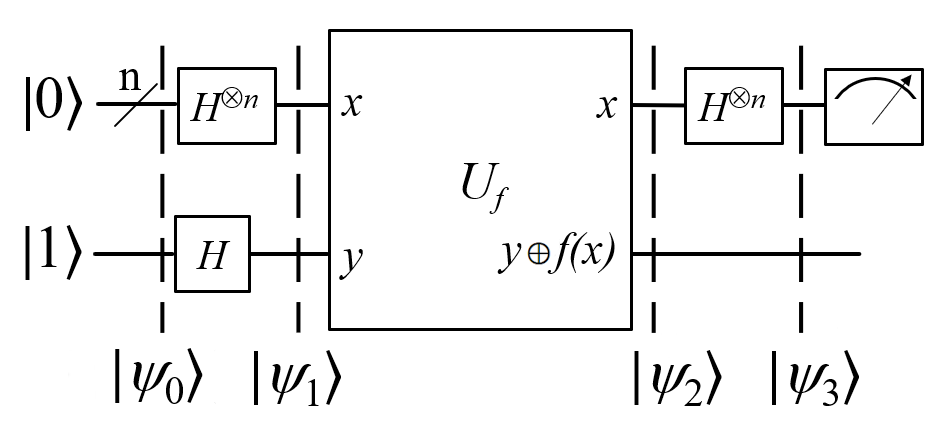}
    \caption{Deutsch-Jozsa quantum circuit to implement the algorithm}
    \label{fig:josza}
\end{figure}

In the worst case, $2^{n-1}+1$ evaluations of \textit{f} would be needed for a standard deterministic algorithm, where n is the number of bits, whereas the Deutsch-Jozsa quantum algorithm can always generate a right outcome, taking a single assessment step.\newline
In Figure \ref{fig:josza} it is shown the circuit to implement the Deutsch–Josza algorithm.

Now we can examine the procedure in more detail, shown in Algorithm \ref{Deutsch-Jozsa}.
\begin{algorithm}
{\small
\begin{enumerate}[labelindent=1cm]
    \item In the first stage, we prepare the input state
    \begin{flalign*}
    |\psi_0\rangle = \bigotimes_{i=0}^{n-1} |0 \rangle\otimes|1\rangle = {|0\rangle}^{\otimes n}|1\rangle = |00...01\rangle &&
    \end{flalign*}
    \item The Hadamard transform is applied to each qubit of $|\psi_0\rangle$
    \begin{flalign*}
        |\psi_1\rangle & = \bigotimes_{i=0}^{n-1} H|0 \rangle\otimes H|1\rangle = \\ &
        =\left({\frac{1}{\sqrt{2}}|0\rangle}+{\frac{1}{\sqrt{2}}|1\rangle}\right)^{\otimes n}\otimes
        \left({\frac{1}{\sqrt{2}}|0\rangle}-{\frac{1}{\sqrt{2}}|1\rangle}\right)=\\ &
        = \frac{1}{\sqrt{2^{n+1}}}\sum_{x=0}^{2^n-1}|x\rangle\left(|0\rangle-|1\rangle\right) &&
    \end{flalign*}
    \item The \textit{f} function has been introduced as a quantum oracle. The oracle converts the state $|x\rangle|y\rangle$ to $|x\rangle|y\oplus f(x)\rangle$ ($\oplus$ is the addition operation in modulo 2 arithmetic). When the quantum oracle is used, we get:
    \begin{flalign*}
        |\psi_2\rangle & = \frac{1}{\sqrt{2^{n+1}}}\sum_{x=0}^{2^n-1}|x\rangle\left(|f(x)\rangle-|1\oplus f(x)\rangle\right) =\\ &
        = \frac{1}{\sqrt{2^{n+1}}}\sum_{x=0}^{2^n-1}(-1)^{f(x)}|x\rangle\left(|0\rangle-|1\rangle\right) &&
    \end{flalign*}
    as $f:\left\{0, 1\right\}^n\rightarrow \left\{0, 1\right\}$ can only get values either 0 or 1.
    \vspace{10pt}
    \item Let $|\widetilde{\psi_2}\rangle$ be such that $|\psi_2\rangle=|\widetilde{\psi_2}\rangle\otimes |y\oplus f(x)\rangle$ and $|\widetilde{\psi_3}\rangle$ such that $|\psi_3\rangle=|\widetilde{\psi_3}\rangle\otimes |y\oplus f(x)\rangle$:
    \begin{flalign*}
        |\widetilde{\psi_2}\rangle & = \frac{1}{\sqrt{2^n}}\sum_{x=0}^{2^n-1}(-1)^{f(x)}|x\rangle\;\wedge\;|\widetilde{\psi_3}\rangle = H^{\otimes n}|\widetilde{\psi_2}\rangle\;\Rightarrow \\ &
        \Rightarrow |\widetilde{\psi_3}\rangle = \frac{1}{2^n}\sum_{x,y=0}^{2^n-1}(-1)^{\left(f(x)-x\cdot y\right)}|y\rangle = \\ & =\frac{1}{2^n}\sum_{y=0}^{2^n-1}\left(\sum_{x=0}^{2^n-1}(-1)^{f(x)}\cdot(-1)^{x\cdot y}\right )|y\rangle
        &&
    \end{flalign*}
    with $x\cdot y = \bigoplus_{i=0}^{n-1}x_i\cdot y_i$\newline
    This means that
    \begin{flalign*}
        \langle\widetilde{\psi_3}\;|\Psi\rangle\langle\Psi |\;\widetilde{\psi_3}\rangle= \left|\frac{1}{\sqrt{2^n}}\sum_{x=0}^{2^n-1}(-1)^{f(x)}\right|^2,\;\;\;\;\;with\;\;|\Psi\rangle=\bigotimes_{i=0}^{n-1}|0\rangle
        &&
    \end{flalign*}
    which is the probability for the state $|\;\widetilde{\psi_3}\rangle$ on the eigenstate $|\Psi\rangle$.
\end{enumerate}
}\caption{Deutsch-Jozsa algorithm.\label{Deutsch-Jozsa}}
\end{algorithm}
Therefore, if \textit{f(x)} is \textit{constant} (\textit{constructive interference}), the probability is 1; if \textit{f(x)} is \textit{balanced}, it evaluates to 0 (\textit{destructive interference}). In other words, if measurement result is the state $|00...00\rangle$ then \textit{f(x)} is \textit{constant}; any other state means \textit{f(x)} is \textit{balanced} and all this in \textit{only one shot}.

\subsubsection*{The Grover's algorithm}\label{Grover}
Grover's algorithm\cite{grover1997quantum}, also known as the quantum search algorithm, is a quantum algorithm for unsorted search on a dataset that uses only $O(\sqrt{N})$ function evaluations to find the unique input to a black box function that generates the correct output value with a high probability; $N=2^n$ is the number of item in our dataset, and \textit{n} is the amount of needed information to represent the whole set. The criterion for the search is implemented in a previously defined function, called \textit{oracle}; this function returns whether or not the input number meets the search criteria.\newline
A common form for this function is
\begin{equation*}
    f(x)=\left\{\begin{matrix}1&if\;\;x=\omega\\0&if\;\;x\neq \omega\end{matrix}\right. 
\end{equation*}
where $\omega$ is the solution for our problem.
\begin{figure}[!ht]
    \centering
    \includegraphics[width=\linewidth]{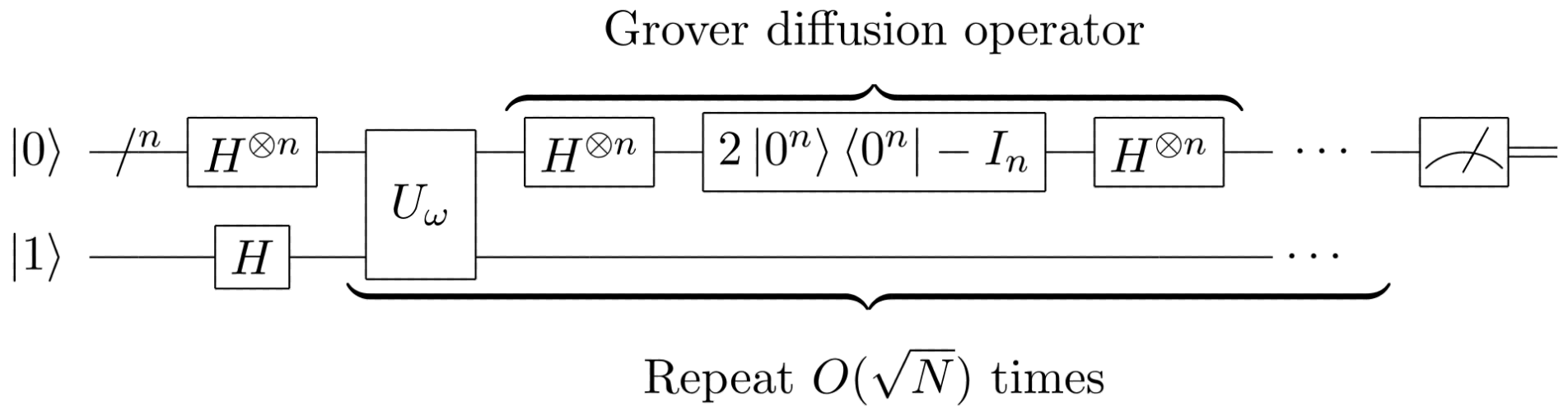}
    \caption{Quantum circuit to implement the Grover's algorithm (Source Wikimedia)}
    \label{fig:grover}
\end{figure}

Now we can examine the procedure in more detail, shown in Algorithm \ref{grover} referring to the Figure \ref{fig:grover}.
\begin{algorithm}
{\small
\begin{enumerate}[labelindent=1cm]
    \item In the first stage, we prepare the input state
    \begin{flalign*}
    |\psi_0\rangle = \bigotimes_{i=0}^{n-1} |0 \rangle\otimes|1\rangle = {|0\rangle}^{\otimes n}|1\rangle = |00...01\rangle &&
    \end{flalign*}
    and then we apply Hadamard transform to each qubit of $|\psi_0\rangle$
    \begin{flalign*}
        |\psi_1\rangle & = \bigotimes_{i=0}^{n-1} H|0 \rangle\otimes H|1\rangle = \\ &
        =\left({\frac{1}{\sqrt{2}}|0\rangle}+{\frac{1}{\sqrt{2}}|1\rangle}\right)^{\otimes n}\otimes
        \left({\frac{1}{\sqrt{2}}|0\rangle}-{\frac{1}{\sqrt{2}}|1\rangle}\right)=\\ &
        = \frac{1}{\sqrt{2^{n+1}}}\sum_{x=0}^{2^n-1}|x\rangle\left(|0\rangle-|1\rangle\right) &&
    \end{flalign*}
    \item The next block, consisting of the oracle and the diffusion operator, must be repeated a number of times less than or equal to $\left \lceil \frac{\pi}{4}\sqrt{N} \right \rceil$; that means algorithm complexity is $O(\sqrt{N})$.\newline
    The oracle will have the task of marking the desired item with a negative phase, leaving the phases of the other elements unchanged.
    \begin{flalign*}
        U_\omega|x\rangle=\left\{\begin{matrix}\;\;\,|x\rangle&if\;\;x=\omega\\-|x\rangle&if\;\;x\neq \omega\end{matrix}\right.\;\;\Leftrightarrow\;\;U_\omega|x\rangle=(-1)^{f(x)}|x\rangle
    \end{flalign*}
    Using a general oracle, it will be possible to trace it back to our particular problem by adding an ancillary qubit with state $H|1\rangle$
    \begin{flalign*}
        & U_f|x\rangle|y\rangle=|x\rangle|y\oplus f(x)\rangle\;\wedge\;
        |y\rangle=H|1\rangle=|-\rangle= \\& ={\frac{1}{\sqrt{2}}|0\rangle}-{\frac{1}{\sqrt{2}}|1\rangle}
        \Rightarrow \\ &
        \Rightarrow U_\omega|x\rangle|-\rangle=(-1)^{f(x)}|x\rangle|-\rangle
        =U_f|x\rangle|-\rangle
    \end{flalign*}
    In the next stage, the diffusion operator will have the task of overturning the counter-phase component back, amplifying its probability amplitude; for the condition of normalization of the state, all the other components will suffer a consequent attenuation. The operator presents the following form:
    \begin{flalign*}
    U_s=H^{\otimes n}\cdot\left(2{|0\rangle}^{\otimes n}{\langle0|}^{\otimes n}-\mathbb{I}_n \right )\cdot H^{\otimes n}
    \end{flalign*}
    Repeating the operation a number of times proportional to $\sqrt{N}$, the probability amplitude for the w component will tend to 1 in absolute value, while all the other components will tend to zero.
\end{enumerate}
}\caption{The Grover algorithm.\label{grover}}
\end{algorithm}

The logic loop can be implemented circuitly by constructing a sequence of Grover operators (oracle and diffusion operators) and repeating them $\sim\sqrt{N}$ times.
The obtained speedup with this algorithm is quadratic.

\subsubsection*{The Shor's algorithm}\label{Shor}
The \textit{general number field sieve} is the most effective classical factorization algorithm known at this time, and it takes exponential times of the order
\begin{equation*}
    O\left(e^{c(logN)^{\frac{1}{3}}(loglogN)^{\frac{2}{3}}}\right)
\end{equation*}
to factor a big enough number N\cite{lenstra1993number, bernstein1993general}.

Shor's quantum algorithm\cite{shor1999polynomial}, on the other hand, could factor an integer in polynomial time; in particular, the complexity of this algorithm is:
\begin{equation*}
    O\left((logN)^{2}(loglogN)(logloglogN)\right)
\end{equation*}

This algorithm did not take long to pique many people's interest due to its potential to break classical cryptography; in fact, the reliability of many cryptographic systems is strongly dependent on the complexity of factoring large numbers; therefore, if a fast and safe method for factoring large numbers was developed, many current cryptographic systems might become vulnerable.

Shor's algorithm solves the following problem: Determine the values of \textit{p} and \textit{q} given a composite number \textit{N = pq} and prime numbers \textit{p} and \textit{q}. The solution to the problem is to return to evaluating the order, or period, of a given function.  

We can examine the procedure in more details, shown in Algorithm \ref{shor}.
\begin{algorithm}
{\small
\begin{enumerate}
\item Choose a random integer number in $1 < a < N$ and use Euclid's algorithm to find the greatest common divisor $gcd(N, a)$.\newline If $gcd(N, a)\neq 1$, a factor of N has been discovered. If, on the other hand, $gcd(N, a)=1$, continue to step 2.
\item 
Determine the order of $a\;mod\;N$, that is, the smallest integer value of \textit{r} such that $a^{r}\;mod\;N=1$.
\item Evaluate \textit{r}
    \begin{enumerate}[label*=\arabic*.]
        \item If \textit{r} is odd, go back to step 1
        \item Else calculate $a^{r/2}\;mod\;N$
    \end{enumerate}
\item Evaluate $a^{r/2}\;mod\;N$
    \begin{enumerate}[label*=\arabic*.]
        \item If  $a^{r/2}\;mod\;N=-1$ go back to step 1
        \item Else continue to step 5
    \end{enumerate}
\item 
Calculate, using Euclid's algorithm, the greatest common divisor
\begin{equation*}
    gcd(a^{r/2} + 1, N)\;,\;gcd(a^{r/2} - 1, N)
\end{equation*}
Both of them are non-trivial factors of $N$.
\end{enumerate}
}\caption{The Shor algorithm.\label{shor}}
\end{algorithm}

Steps 1 and 3 can also be performed efficiently by a classical computer, while a quantum computer can more easily find the period of the function \textit{f} in step 2, thus obtaining the required speed-up compared to the classic computer.
Shor's algorithm could be used to crack public-key cryptographic schemes like the commonly used RSA scheme if a quantum computer with enough qubits could operate without succumbing to quantum noise and other quantum decoherence phenomena.
The commonly used RSA asymmetric encryption algorithm, developed in 1977 by Ronald Rivest, Adi Shamir, and Leonard Adleman, is based on the principle that factoring large integers is computationally intractable. This theorem holds true for classical computers: no classical algorithm that can factor integers in polynomial time is known. However, Shor's algorithm demonstrates that the factorization of integers is effective on an ideal quantum computer; hence, it could potentially be able to breach RSA if we had quantum computers with a large number of qubits, which do not yet exist in the current state of technology.

This does not rule out the possibility of implementing this algorithm on a real quantum computer in the future: the only way to protect against quantum computer threats is to develop modern cryptographic schemes.

\subsection{Quantum Computing insights}
After years of solid theoretical work to appropriately approach the domain of quantum computation, despite several breakthroughs in the fields of nanoelectronics and photonics, quantum computer that is reliable, scalable, stable, resilient to environmental noise, and general-purpose has not yet attained maturity.
Indeed, as previously said, quantum computing is a very new field; we are thus in a historical period quite comparable to that of the first explorers of the New World.


We tend to think of computers as machines capable of manipulating enormous amounts of data in a reasonable time, beyond the semantics of the data itself; but they remain fundamentally and intimately fast logic-arithmetic computing machines. 
Furthermore, their peculiar architecture gives them a great lot of flexibility in dealing with a wide range of challenges.
We will here ignore quantum computers with a large number of qubits that are dedicated by nature to solving specific optimisation problems (adiabatic quantum computers), such as D-Wave \cite{gibney2017d}, and instead focus on general-purpose quantum computers.
The current metric that allows to classify a quantum computer as general-purpose is based on Di Vincenzo's five criteria \cite{divincenzo2000physical}; they state that a quantum computer should:
\begin{enumerate}
    \item be feasible and physically realized through a scalable technology and in which the qubits are easy to be identified and not too difficult to be managed;
    \item be able to return to a predetermined fiducial state; typically, the $\bigotimes_{j=1}^n |0\rangle$ state is chosen;
    \item have long enough decoherence periods to allow the single quantum gates that make up the circuit to carry out the transformation operations; in other words, qubit decoherence periods have to be longer than gate action times;
    \item possess an universal base set of gates, so that any sequence of unitary transformations may be performed;
    \item be able to undertake a precise and defined qubit measurement process that permits it to be collapsed onto one of the computational base's components.
\end{enumerate}
Once we have become aware of the scenario in which we are moving, it might be interesting to experiment with the theoretical knowledge acquired through the numerous tools made available by great IT giants, public and private institutions, and individual enthusiasts.

Many of them provide a variety of analysis and synthesis tools for developing a quantum algorithm; some are sophisticated simulators, while others are cloud services that, after registration, allow you to immediately access a real quantum computer.
While the simulators automatically adapt to any form of scheme by employing very efficient compilers that implement various optimisations on the suggested code, the use of a real quantum computer for very sophisticated and elaborate algorithms surely requires an expert eye.

The main frameworks and/or programming environments for exploring quantum architectures are now presented.
\subsubsection*{IBM's QExperience and Qiskit} 
It is an online platform that enables quantum computation on quantum architectures connected to the Cloud.
However, before implementing the generated circuit, you must select which backend to use among those offered.
The project's execution will be queued with the other tasks requested by other users.
\begin{figure}[!ht]
    \centering
    \includegraphics[width=\linewidth]{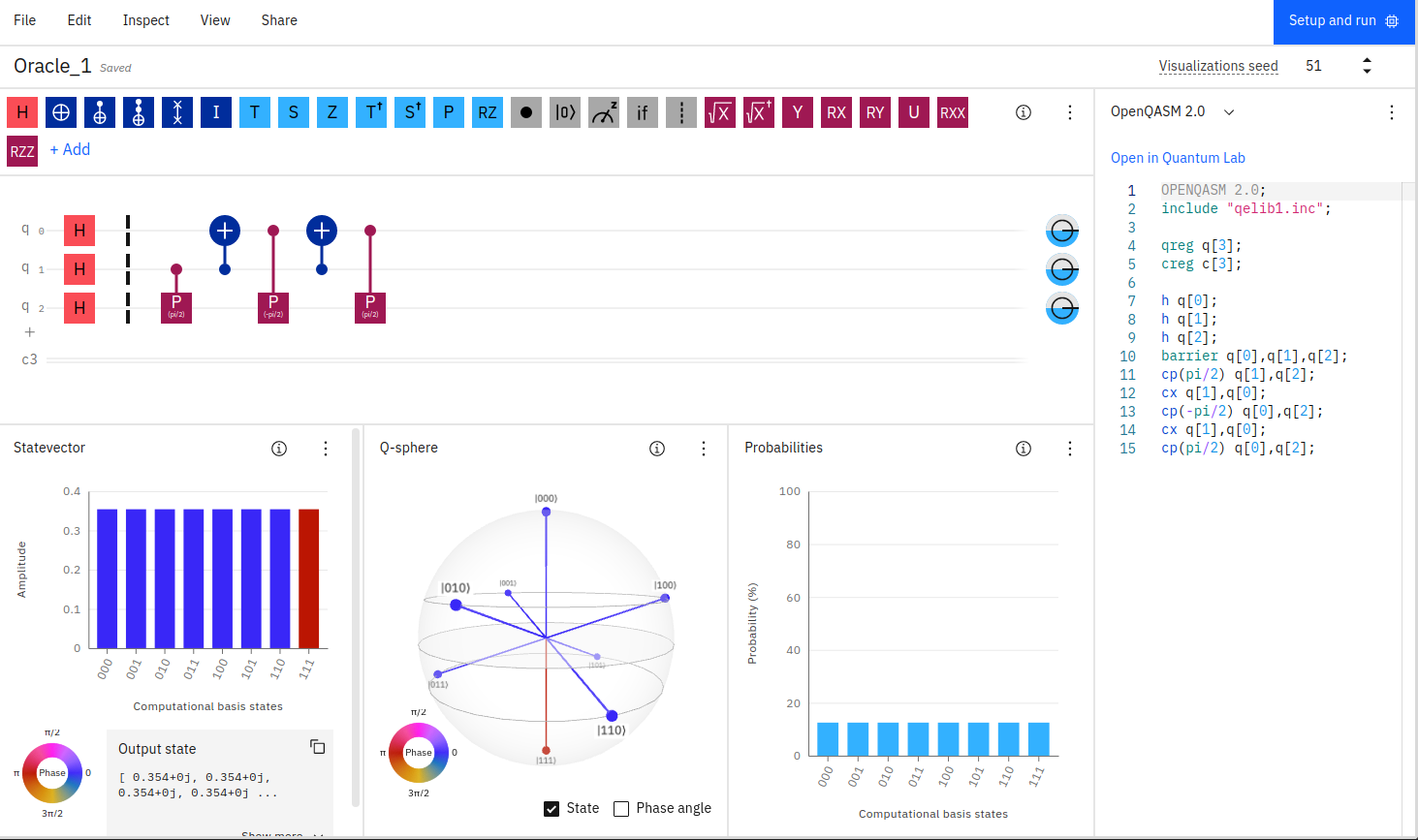}
    \caption{IBM quantum composer example}
    \label{fig:ibmQexp}
\end{figure}
It is also possible to instantiatly run multiple simulations by downloading a dedicated Python library, initially released by IBM but now Open-Source, called Qiskit\footnote{Qiskit official website: \url{https://quantum-computing.ibm.com/}, \url{https://qiskit.org/}}. 
Programming is very intuitive and takes place thanks to a tool called Composer, a fully graphical tool, consisting of an area of five lines, each reserved for a qubit, as shown in Figure \ref{fig:ibmQexp}.
Obviously, quantum logical operators can be inserted within the composer, which can be selected from a "palette" and placed in our circuits through the action of Drag and Drop; at the end of each line there must be the operator that allows the measurement of the quantum state of the qubit corresponding to the line, storing the result on a classical bit representation. 
The results, once obtained, will be represented by means of a histogram. 
IBM Quantum Experience also provides a text-level programming mode that takes advantage of the QASM language, very similar to C.

IBM provides users with five quantum simulators with different characteristics.
The first is the Clifford simulator (stabiliser), which makes it possible to simulate Clifford circuits with and without noise.
The second is the Matrix Product State (MPS), which simulates tensor networks.
The third is the Extended Clifford (extended\_stabilizer), which extends the functionality of the stabiliser simulator.
The Extended Stabilizer technique is divided into two main sections. The first is a method for breaking down quantum circuits into stabilizer circuits, a type of circuit that can be efficiently replicated classically. The second is a method of integrating these circuits in order to conduct measurements.
The fourth is the general purpose (Qasm\_simulator), which allows general purpose simulations to be carried out and is the backend that is provided by default for theoretical simulation.
The fifth is the Schrödinger wavefunction (statevector), which simulates a quantum circuit by calculating the wavefunction of the state vector

IBM also provides 20 real quantum computers, of which seven are free, and 13 are chargeable.
These computers can be grouped into various categories according to the technology used by the quantum processors:
Canary, Falcon, Hummingbird.
The Canary family consists of tiny designs having 5 to 16 qubits.
All qubits and readout resonators are on the same layer of the optimized 2D lattice.
The Falcon family of devices provides a key platform for medium-scale circuits, as well as a good platform for proving performance and scalability enhancements prior to pushing them onto bigger devices.
The Hummingbird is a new series of quantum processor that supports up to 65 qubits using a heavy-hexagonal qubit architecture.

\subsubsection*{Google’s Cirq and Tensorflow Quantum} 
Cirq is a Python software library that allows us to writing, manipulating, and optimizing quantum circuits and to run them on quantum computers and simulators. 
Cirq\footnote{Cirq official website: \url{https://quantumai.google/cirq}} tries to highlight the specific characteristics of the hardware, since in the Noisy Intermediate-Scale Quantum regime (also known as NISQ) these peculiarities determine the possibility or not of running a simulation on a given circuit.

Research in quantum machine learning examines the interaction between quantum computing and machine learning.
According to current thinking, quantum computers can be utilised and trained in the same way that neural networks are used.
Physical control factors such as electromagnetic field strength or laser pulse frequency can be adjusted in an organised manner in order to address a problem.
TensorFlow Quantum (TFQ)\footnote{TensorFlow Quantum official website: \url{https://www.tensorflow.org/quantum}} is a quantum machine learning library enabling quick prototyping of hybrid quantum-classical ML models.
Cirq and TensorFlow Quantum are both extremely versatile and feature-rich, and are in great demand. They are also intuitive and simple to use.

\subsubsection*{Xanadu’s Strawberry Fields and Pennylane} 
The open-source software PennyLane\footnote{PennyLane official website: \url{https://pennylane.ai/}} is based on the notion of quantum differentiable programming and is free to download and use. 
Classical machine-learning libraries are easily integrated with quantum simulators and hardware, allowing the user to train quantum circuits.
PennyLane's primary task is to manage the assessment of parametrized quantum circuits (also known as variational circuits) on quantum devices and to make them accessible to machine learning libraries. PennyLane also gives access to quantum circuit gradients, which the machine learning library can utilise to execute backpropagation, including through quantum circuits an important process in optimisation and machine learning.
One promising technology for physically realising a quantum computer is photonic processors.
This technology uses the modes of photons (qumodes) to store information.
Many companies are working on this new type of machine.
Xanadu makes a programming environment called Strawberry fields available to users after registering on the website.

\subsubsection*{Microsoft’s Q\# and Azure Quantum} 
Microsoft Azure\footnote{Microsoft’s Q\# official website: \url{https://azure.microsoft.com/en-us/services/quantum/}} is a cloud-based service that developers, academics, and businesses may utilise to execute quantum computing applications or solve optimisation issues. Q\#, Microsoft's open source programming language for building quantum algorithms, is included in the Microsoft Quantum Development Kit (QDK). It also includes Q\# libraries, Python and .NET APIs, and a Python SDK for optimisation solvers and quantum simulators.
The QDK is a full-featured development kit for Q\# that can be used in conjunction with conventional tools and languages to create quantum applications that can be executed in a variety of settings. Q\# applications can be executed as a console application, in Jupyter Notebooks, or through a Python or .NET host program.
The Q\# libraries allow you to do sophisticated quantum operations without the need to create low-level operation sequences.

The QDK provides numerous quantum machine classes, which are all specified under the namespace Microsoft.Quantum.Simulation.Simulators.
The first is the "Full State Simulator",which replicates a quantum machine on your local computer.
The full state simulator may be used to execute and debug quantum algorithms developed in Q\# and allow you to use up to 30 qubits.
The second one is the "Simple Resources Estimator", it can estimate resources for Q\# operations that utilise thousands of qubits, as long as the classical component of the code executes in an acceptable amount of time.
The third one is the "Trace-based resource estimator" which executes a quantum program without really mimicking a quantum computer's state. 
As a result, the quantum trace simulator can run quantum algorithms with thousands of qubits. 
It help the developer to debug classical code that is a component of a quantum program, moreover calculating the resources needed to run a specific instance of a quantum program on a quantum computer, the trace simulator serves as the foundation for the Resources estimator, which gives a more limited set of indicators.
The fourth one is the "Toffoli simulator", which is a limited-scope special-purpose simulator that only supports X, CNOT, and multi-controlled X quantum operations. 
There is access to all classical logic and calculations; it is not so powerful and effective as the Full State Simulator.

\subsubsection*{Amazon’s AWS Braket} 
According to the AWS\footnote{AWS bracker official website: \url{https://aws.amazon.com/braket/}}  concept, Amazon Braket is a fully managed quantum computing service. 
Amazon Braket provides a development environment for experimenting  and developing quantum algorithms, as well as testing them on quantum circuit simulators and running them on different quantum hardware technologies. 
To design quantum algorithms and manage experiments, it may be used his/her own development environment or fully managed Jupyter notebooks in Amazon Braket.
There are four circuit simulators available with Amazon Braket to perform and evaluate quantum algorithms.
The first one is a free local simulator that may be used on your laptop or on an Amazon Braket managed notebook. Depending on your hardware, the local simulator is well suited for conducting small and medium-scale simulations up to 25 qubits without noise or up to 12 qubits with noise. 
The second one is SV1, a full managed \textit{State Vector} simulator, which calculates the outcome by taking the complete wave function of the quantum state and applying the circuit's operations. 
You can use SV1 to test your quantum method at scale after designing it in the Amazon Braket SDK using the local simulator, and then utilising SV1's parallel processing to perform several batches of simulations.
SV1 can handle simulations of up to 34 qubits.
The third one is DM1, is a fully managed and high-performance simulator that simulates the effects of noise on quantum circuits using \textit{Density Matrix} computations. 
You may define realistic error models for your circuits and investigate the influence of noise on your algorithms using DM1. 
DM1 allows you to simulate circuits with up to 17 qubits.
The last one is TN1 a \textit{Tensor Network} simulator for structured quantum circuits is a fully managed, high-performance tensor network simulator. 
A tensor network simulator converts quantum circuits into a structured graph in order to determine the most efficient approach to calculate the circuit's result.
You can use TN1 to simulate certain types of quantum circuits up to 50 qubits in size.

\subsubsection*{Rigetti’s Forest}
Forest\footnote{Righetti forest official website: \url{https://www.rigetti.com/}} is a cloud computing platform that provides developers with access to quantum processors so they may test quantum algorithms and simulate them by using a quantum device with more than 32 qubits. It is built on a specialised instruction language known as Quil (Quantum Instruction Language), which can enable a form of hybrid computing (simultaneous use of quantum components and classical logic).
Moreover, applications may be developed and run using free and open source Python tools.

\subsubsection*{Quantum Inspire} 
QuTech's Quantum Inspire (QI)\footnote{QuTech. (2020). Quantum Inspire Home. Retrieved from Quantum Inspire: \url{https://www.quantum-inspire.com/}} is a quantum computing platform. The objective of Quantum Inspire is to give users access to various tools for doing quantum calculations, as well as insights into quantum computing concepts and access to the community.
QuTech, the advanced research center for quantum computing and quantum internet created by TU Delft and TNO, started Quantum Inspire (QI).
Quantum Inspire gives users a range of options for programming quantum algorithms, running them, and analyzing the results.
It gives you a graphical interface for programming in QASM (Quantum Assembly Language) and visualizing operations in circuit diagrams.
Non-quantum specialists may learn to write quantum algorithms with the QI Editor, which includes automated bug detection and autocomplete.
Quantum Inspire has three account types: Anonymous, Basic, and Advanced.
You can create some simple algorithms and utilize up to 5 qubits in the QX simulator as an anonymous user.
A basic account allows you to simulate quantum algorithms with up to 26 qubits and run them on our hardware back-ends.
With an advanced account, you may simulate up to 31 qubits on Cartesius, one of the Dutch supercomputers at SURFsara, utilizing more powerful computing resources.

\subsubsection*{Quirk}
Quirk\footnote{A live version of Quirk may be found at \url{https://algassert.com/quirk}, but you can also obtain the source code from \url{https://www.github.com/Strilanc/Quirk} and create your own.} is a quantum circuit simulator that allows you to manipulate and explore tiny quantum circuits by dragging and dropping components.
The visual design of Quirk offers a very intuitive sense of what is going on, state displays update in real time as you modify the circuit, and the whole experience is quick and engaging.
Using Quirk consists primarily of dragging gates from the toolboxes, putting them into the circuit, and inspecting the state displays within and to the right of the circuit.
As you change the circuit, Quirk actively rewrites the URL in the address bar to refer to the current circuit. You may also save the circuit by clicking the Export button above the circuit editing box. You can export an escaping link, an offline copy of Quirk defaulting to the current circuit, a JSON representation of the current circuit, or a JSON representation of the full simulator output.
Quirk is a free and open source software application.
The source code is accessible under a permissive Apache license, allowing anybody to create and distribute updated versions.

\section{Techniques enabling the solution of complex problems based on Computational Intelligence}\label{computationalIntelligence}

\subsection{Approaches based on Machine Learning}\label{neuralNet}

\begin{figure}[!ht]
    \centering
    \includegraphics[width=\linewidth]{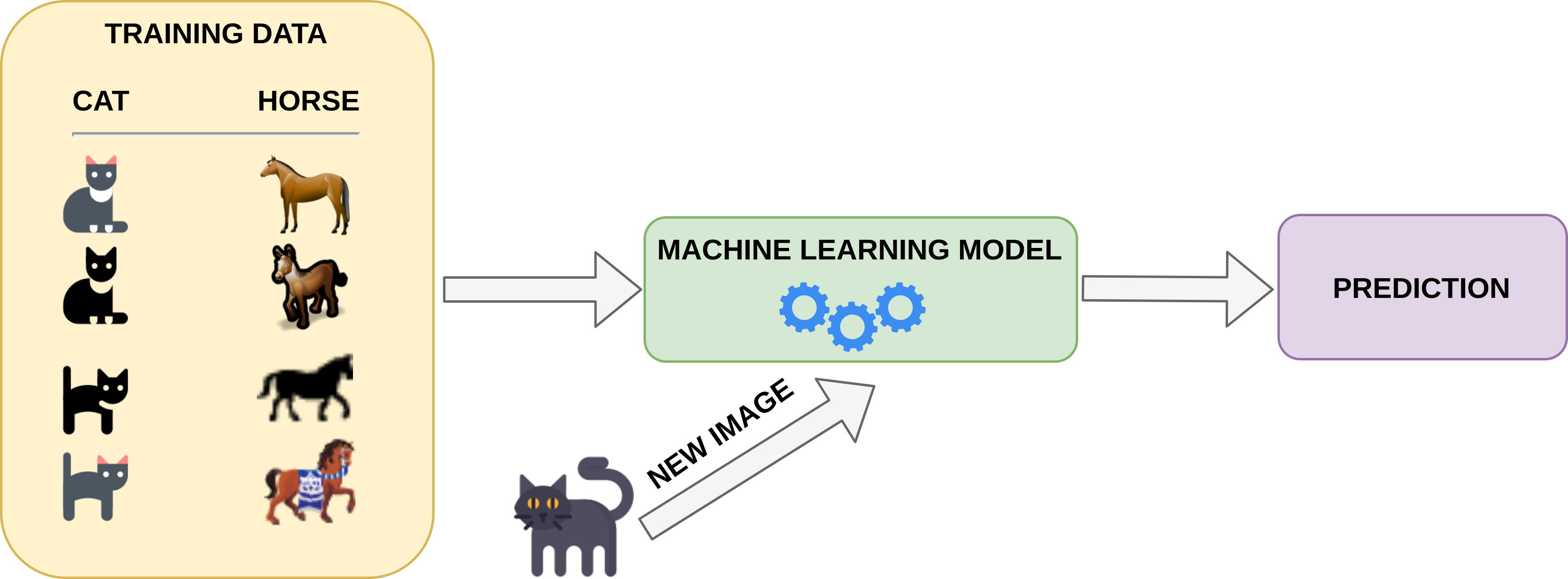}
    \caption{Supervised learning}
    \label{fig:supervisedML}
\end{figure}

There are several types of Machine Learning approaches.
Supervised Machine Learning is one of the most prevalent.
This kind entails creating a dataset of items with which to train the classifier.
The dataset must be accurately constructed, and the items must be classified into classes.
The dataset is then split into at least two parts: training and testing.
The training set is used to train the classifier, while the test set is used to validate the results.
The classifier is then trained and should learn to recognize the input items.
Following the completion of the training, the quality of the classifier created is evaluated by presenting images that were not part of the dataset utilized during the training process.
An example is shown in Figure \ref{fig:supervisedML}.

The following are some of the most often exploited machine learning techniques today.

\subsubsection*{Decision trees}
Decision trees (DT) are a non-parametric, supervised machine learning method for classification and regression.
DTs generate "white box" models that are easily interpretable as if-then-else expressions. Small data modifications, on the other hand, might result in the generation of wholly distinct trees.

\subsubsection*{Random Forest}
Random Forest fits many independent trees on various subsets of the dataset to reduce variance and over-fitting.
Gradient Tree Boosting is another notable tree-based classifier, which uses gradient descent to gradually grow a weighted ensemble of DTs.

\subsubsection*{Bayesian classifier}
The simplest variant of a Bayesian network is the naive Bayesian classifier (NBC).
Given the class variable, an NBC assumes that all features are conditionally independent.
While this assumption is frequently incorrect, it has proven to be extremely effective in practise, obtaining good, simple models with little training.

\subsubsection*{Logistic Regression}
Logistic Regression (LoR) is a simple type of regression analysis in which the independent variables and the log-odds of the classes are assumed to have a linear relationship.

\subsubsection*{K-Nearest Neighbours}
K-Nearest Neighbours(KNN) finds the K data points closest to the query feature vector and polls their assigned labels to determine the query vector's label.
KNN is a non-generalizing method, that means instead of learning a model's parameters, it 'remembers' the training points and eventually stores them in a suitable data structure, such as a Ball Tree (i.e. metric tree), to speed up inference.

\subsubsection*{Support Vector Machine}
Another machine learning technique is the Support Vector Machine.
The data is separated using a support vector machine (SVM) that draws hyper-planes in the feature space. To maximise the separation between classes, the points closest to the hyper-planes are used as a reference. The 'kernel trick', which entails defining alternative inner products for the data points, can be used to implicitly embed the problem in higher-dimensional spaces, allowing for complex separating surfaces.

\subsubsection*{Multi Layer Perceptron}
The simplest feed-forward neural network is the Multi Layer Perceptron (MLP). 
There are at least three node layers in it: an input layer, a hidden layer, and an output layer. Each node, with the exception of the input nodes, uses a nonlinear activation function.

\subsubsection*{Convolutional neural network}
CNNs are a type of deep, feed-forward artificial neural network that is commonly used for image recognition.
They are made up of a series of layers, each of which applies a set of learned convolution filters to the previous layer's results (or 'activations').
The final layer is usually a classifier with a loss function to back-propagate gradients through the network and update the filter values ('weights').
Non-linearity is typically added after each convolution layer, as well as other layers like pooling operators and fully-connected layers.
CNNs convert an input image's original pixel values into final class scores.

\subsection{Machine learnings insights}
Galileo Galilei (1564-1642) introduced the scientific method, characterising and shaping science for centuries to come.
The scientific method is based on the repeatability of experiments, which ensures that researchers and scientists can verify the validity of results produced by others.
Machine learning, and in particular neural network learning, on the other hand, produces different results each time learning is carried out.
This is due to the fact that the weights of the neurons that make up the neural network and that are to be trained are instantiated at random values and the search for values that optimise the objective function can lead to different search paths at each iteration.\cite{9266043} 
Because of these characteristics, machine learning makes it possible to address issues, studies and research in a completely new way.
For example, in recent times our research group has started a study that seeks to achieve protein recognition through the use of machine learning.
We started with a method based on the classification of proteins by analysing the list of proteinogenic amino acids that make them up.
We analysed this list using a one-dimensional convolutional neural network capable of discriminating a real protein from a fake one.\cite{DBLP:conf/iccsa/PerriSLLG20}
Subsequently, we tried to improve the work carried out and analysed the three-dimensional structure of the proteinogenic amino acid chain by including the list of X,Y,Z coordinates of the various elements in our analysis.\cite{iccsa/protein2021impress}
This type of analysis thus produced a four-dimensional matrix, since each element is composed of the 3 spatial coordinates and the type of amino acid, coded with an integer.
Since we wanted to test the quality of the analysis by means of a convolutional neural network, we were faced with a problem: most neural networks used in the literature accept three-dimensional matrices as input.
We therefore devised a method to transform 4D-encoded proteins into 3D-encoded proteins.
We have implemented a graphic representation based on orthogonal axonometry.
\begin{figure}[!ht]
    \centering
    \includegraphics[width=0.48\textwidth]{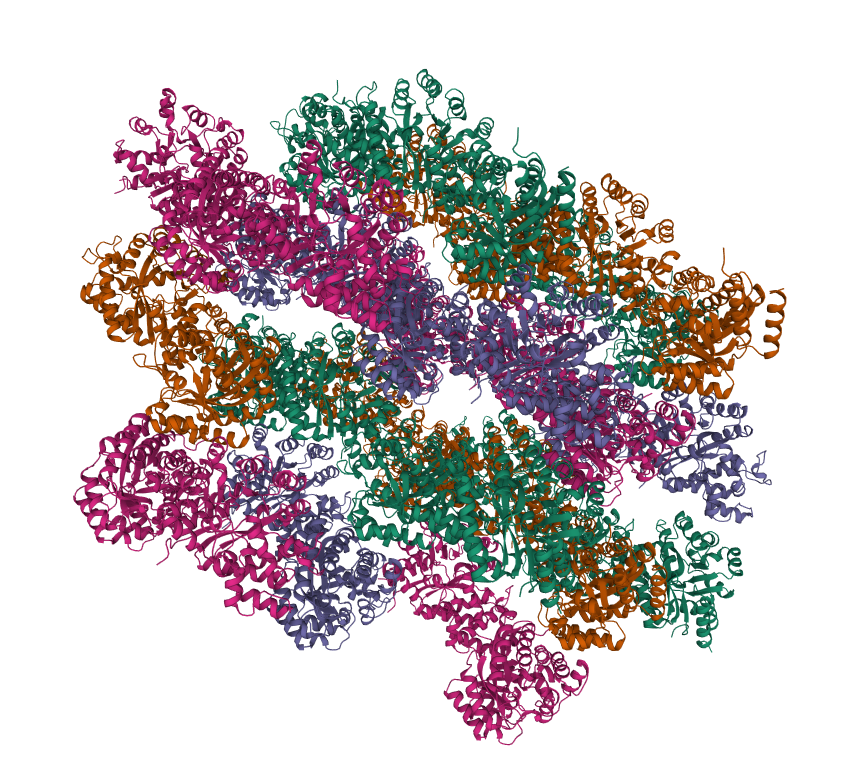}
    \includegraphics[width=0.48\textwidth]{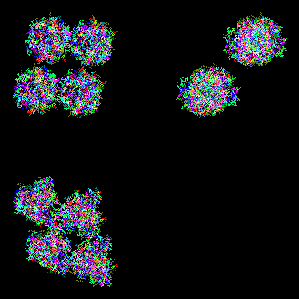}
    \caption{Ada Structure Complexed With Deoxycoformycin from PDB (left hand side), Orthogonal Axonometry representation (right hand side)}
    \label{fig:proteinPDB}
\end{figure}
We divided an initially empty image into 4 parts and inserted the proteins, encoding the colour of each pixel with the type of amino acid, and the XY position using a transformation function that takes photographs of the protein when viewed from 3 different directions.
In a single image we have therefore inserted a protein, encoded by its representation in cavalier axonometry.
We then analysed the images using convolutional neural networks and obtained a higher performance than in the previous work.
In Figure \ref{fig:proteinPDB} is shown an example, considering the 3D representation of the \textit{Ada Structure Complexed With Deoxycoformycin} provided by the Protein Data Bank (PDB) (left hand side) and our representation using the Orthogonal Axonometry (right hand side).

The programming of applications that make use of machine learning is greatly facilitated by a number of frameworks that allow developers to implement high-level algorithms that benefit the computational capabilities of machine learning approaches and that use a number of already optimised libraries.
In Table \ref{tab:frameworks} are summarised the main Frameworks available.
\begin{table}[]
\centering
\resizebox{\textwidth}{!}{%
\begin{tabular}{|p{0.20\textwidth}|p{0.47\textwidth}|p{0.32\textwidth}|}
\hline
\textbf{Framework} & \textbf{Main characteristics} & \textbf{Released by} \\ \hline
Keras & Open Source (MIT License), written in Python & François Chollet \\ \hline
Scikit-learn & Open Source (BSD-3-Clause License), written in Python, Cython, C and C++ & David Cournapeau \\ \hline
Tensorflow & Open Source (Apache2 License), written in C++ and CUDA & Google Brain Team \\ \hline
pyTorch & Open Source (BSD License), written in Python, C++ and CUDA & Facebook's AI Research lab (FAIR) \\ \hline
Microsoft Cognitive Toolkit & Open Source (MIT License), written in C++ & Microsoft Research \\ \hline
H2O & Open Source (Apache2 License), written in Java & H20.ai \\ \hline
\end{tabular}%
}
\caption{Main frameworks for machine learning}
\label{tab:frameworks}
\end{table}

\section{The dilemma of respecting privacy in multichaos situations}\label{multichaos}
The subject of personal data and the investigation of methods and policies capable of safeguarding them has been one of the most contentious concerns in recent years.
Most machine learning approaches make extensive use of enormous volumes of input data, which analysts and data scientists use to train and calibrate their computer systems.
In most circumstances, both neural networks and traditional classifiers require thousands of samples to train their classification and prediction algorithms.
These technologies are now widely employed in a variety of sectors.
So, it is important to pay great attention to which data can and cannot be utilized.

\subsection{GDPR}\label{GDPRsubsubsection}
The EU General Data Protection Regulation (GDPR) was issued on May 25, 2018, and it replaced the Data Protection Directive, which had been in effect since 1995.

\begin{figure}[!ht]
    \centering
    \includegraphics[width=\linewidth]{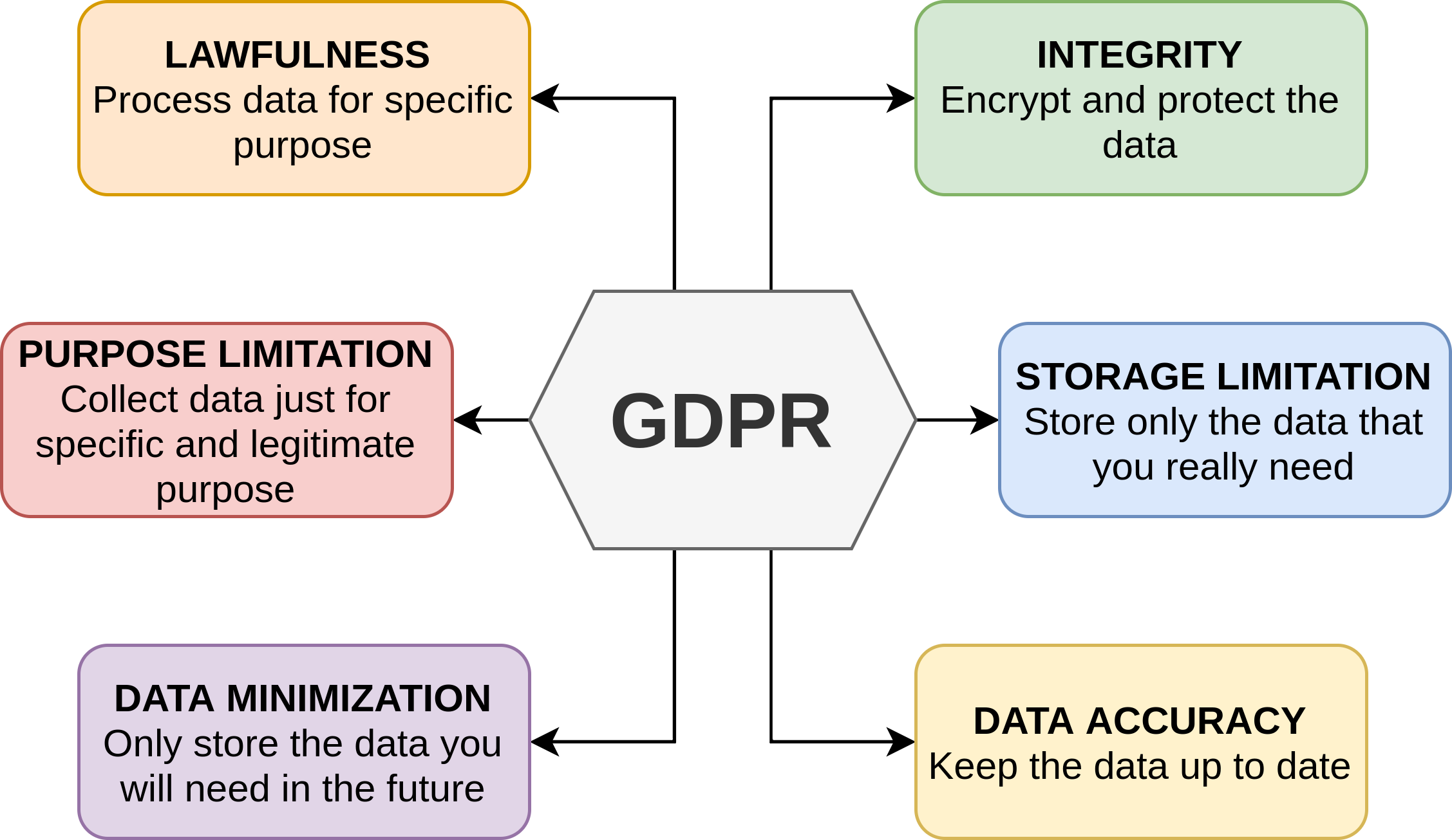}
    \caption{GDPR 6 principles}
    \label{fig:gdpr}
\end{figure}

The GDPR \cite{GDPR2020effects,gdprbusiness,gdrpGlobalImpact} is described as a contemporary and effective instrument aimed at ensuring user privacy and protecting them from illicit use in a historical context that highlights a vast usage of personal or sensitive data through information technologies.
The rule is founded on six principles, as seen in the Figure \ref{fig:gdpr}.
If we are dealing with data obtained from users of our services, we must guarantee that they are informed of the profiling through a written document that describes the treatment that will be done and the information that will be gathered in a clear, intelligible, and effective manner.
DPA stands for Data Processing Agreement, and it is a document required by Article 28 of the GDPR.
The organisation that controls the data must create and publish this document, which must clearly clarify the circumstances and procedures of processing users' personal data.
It should also be stated what sort of data will be held, how long it will be retained, and how its security will be assured.

It is also vital to apply all best practices to guarantee that the data is maintained securely and is protected from cyber assaults that might facilitate its disclosure for illicit reasons.

To accomplish the desired result, the prior data protection model is greatly enlarged, and the relationships, duties, and powers of the different national and international subjects are explicitly described.
This new set of rules and regulations applies within the European context and is implemented for both business with internal offices in the European Union and external enterprises which exchange products or services with the European Union.

Based on Article 37, paragraph 7 of the Regulation, the GDPR indicates the presence of a new legal figure: the Data Protection Officer (DPO).
The DPO is responsible for ensuring that the processing of personal data conforms with the law and the DPA.
The statute specifies the situations in which the designation of the DPO is required.

These legal entities are wholly new figures who contribute to enhance and make the concept of accountability effective, so increasing the degree of protection of users' privacy and, hopefully, their trust in the digital world over time.

\subsection{AI and privacy}\label{AIsubsubsection}
In recent months, the European Union has been debating how to regulate artificial intelligence, its applications, and its interference with people's privacy and life.
The proposal was made on April 21, 2021 (2021/0106), and it contains various goals.
First and foremost, it is necessary to make certain that artificial intelligence applied within European borders does not infringe laws or basic rights.
The usage of these technologies is also promoted, making investments and innovative procedures easier.
The purpose is to strengthen and improve existing legislation by specifying the minimal safety and dependability standards for artificial intelligence systems.

The intent is to make it easier to create a unified market for law-abiding AI applications.
\begin{figure}[!ht]
    \centering
    \includegraphics[width=\linewidth]{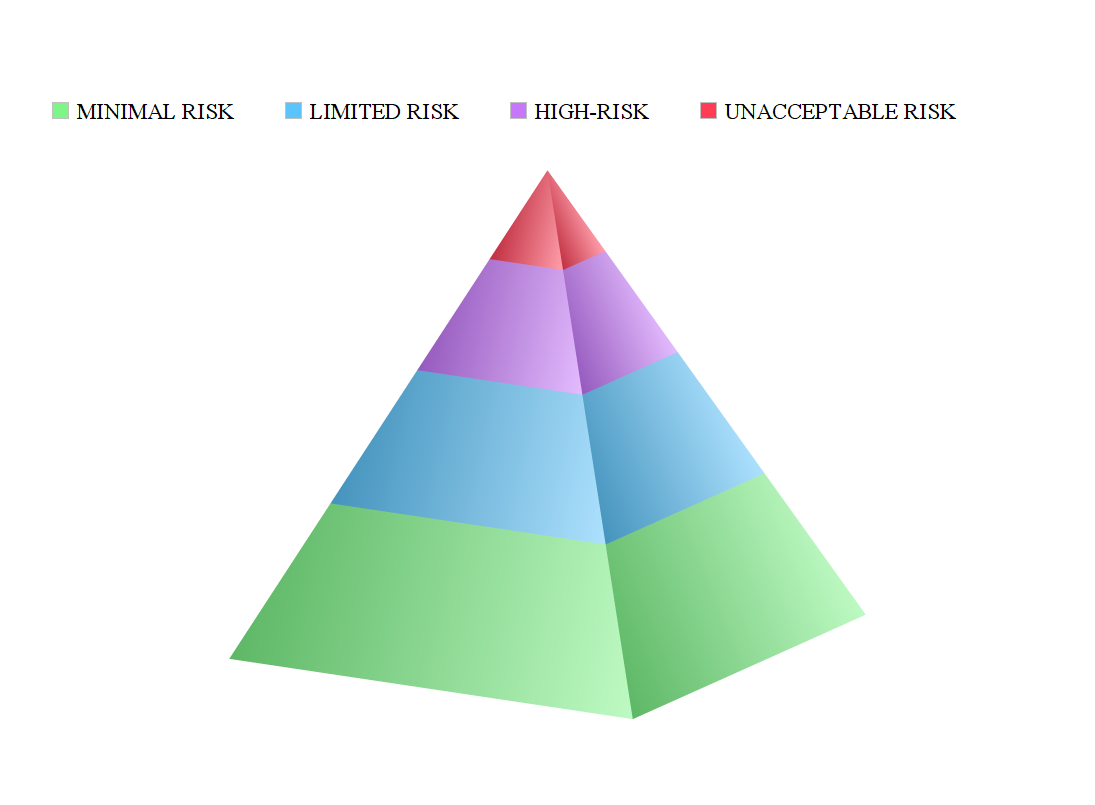}
    \caption{Classification of artificial intelligence algorithms according to the risk associated with their usage}
    \label{fig:aiclassification}
\end{figure}

Some illegal applications of artificial intelligence are defined in the proposed rule.
It is important to prevent distorting the impression of reality that the most vulnerable individuals may undergo because of the employment of subliminal messaging.
It is illegal to create artificial intelligence algorithms that affect a person's behavior based on their age or handicap.
Another restriction pertains to social scoring.
As a result, the European Union seeks to prohibit bulk analysis of groups of individuals, as it is commonly done in some countries in the world.
Somewhere, the social credit score is calculated by continuously evaluating user behaviors and activities with machine learning algorithms.

For example, a person who smokes a cigarette in a no-smoking area or fails to obey a red light receives a lower score.
The European initiative intends to outlaw social scoring and associated practices.
Finally, it is intended to prohibit the use of biometric data-based systems for identifying persons, such as face recognition.
The latter restriction would apply in public places and for the sake of crime suppression.

There are intelligent tracking technologies that help a user to obtain various types of information on locations or objects without requiring the use of the user's personal and sensitive data\cite{DBLP:conf/iccsa/GervasiFMPS19}.

Artificial intelligence applications are also divided into four categories as shown in Figure \ref{fig:aiclassification}.
The first group is labeled "minimal risk," and it includes things like spam filters in emails and artificial intelligence in video games.

Chatbots\footnote{Chatbots are software capable to interact with humans and they are able to answer preset and simple questions} are an example of a product in the second category, which is referred to as "limited risk."
The rule states that users must be informed of their engagement with a machine and must be free to choose whether or not to continue the engagement.
The third classification is "High-risk."
This area comprises artificial intelligence systems for legal proceedings, document authentication, judging student examinations, managing public or private transportation, and so on.

The fourth category is titled "Unacceptable risk," and it includes AI that is deemed a risk to society.
Social scoring or AI-based toys with voice assistants are two examples.

To make compliance with the new legislation easier, "sandboxes" are being considered, which will aid in the creation of artificial intelligence in order to produce "legal by design" applications.
Some European countries have already implemented them, and the proposed European rule will formalise them.

A European committee on artificial intelligence is also being considered.
This new organisation, which shares some of the tasks of the European Data Protection Committee (EDPB), is in charge of verifying and supervising the execution of the Regulation inside the European Union's member states, as well as drafting the new rules.

The new European legislation represents the first attempt to regulate the use of artificial intelligence-based technologies so that they can be used to enhance the efficiency of various applications in different contexts, without causing harm or detriment to users and organisations.

\section{Conclusions}\label{conclusions}
In this paper we have addressed various topics that help researchers and developers to deal with complex and chaotic situations.
We think that such situations need a massive conceptual effort to comprehend and explain their features and the most sophisticated technology to achieve the specified goals in the shortest time and with the greatest precision.
The technologies we are describing, accompanied by some guidelines that can facilitate their implementation, are essential in order to face the challenges of modern times.
If these technologies are early adopted by some organisations, they can be helpful in the way the organisation deals with the complex situation.

Thanks to containers, virtualised environments can be realised in a highly efficient way that can even be recycled for future projects with little effort.
Until a few years ago, it was common to build virtual machines for specific services, which took up a lot of resources in terms of both computation and disk space.
Today, however, a web container can occupy as little as 50MB of RAM in idle state and keep only the very few processes required to deliver the service active.

We have also presented how neural networks are increasingly used in research and how they are effectively accelerated by programming techniques that run on graphic cards. 
Modern GPUs are, in fact, auxiliary devices that perform general purpose calculations alongside modern CPUs.
Thanks to neural networks and deep learning techniques, it is possible, for example, to solve very complex problems such as the recognition of emotions, the recognition of clinical pathologies by means of images, etc.
Neural networks are also used in robotics, mechanics, manufacturing and electronics (e.g. to recognise faulty chips or predict failures and malfunctions), the automotive industry and other sectors.

The growing demand to tackle computationally hard issues has fueled a frenzied race to construct actual quantum computers: a gold rush that began just over 20 years ago from a scientific standpoint, but is now witnessing the fall into the field of major technical titans, who aim to monetise investment incentives in the area over the next decade. 
From the standpoint of science and study, quantum computing would be a significant step forward, allowing us to open the door to fresh perspectives on our knowledge of the world around us.

Machine learning is a branch of scientific research that has had a strong wind of renewal since the 2000s and that today allows the use in the real world of theoretical codes and algorithms that exploit the computational capabilities of modern hardware.
Furthermore Machine learning is of great help to professionals and experts in making decisions, classifying images or predicting the behaviour of certain complex phenomena such as meteorology or financial trends.
This discipline is extraordinarily important when experts are faced with complex situations as it develops approaches and provides solutions in less computation time than the conventional approach.

Finally, we addressed the issue of privacy.
There are two important European regulations: the GDPR which is approved and operational, and the European regulation on artificial intelligence which is under development.
These regulations impose a whole series of constraints and obligations to respect the freedom, privacy and well-being of the European population.

  \appendix
\section*{Acronyms}
The following acronyms are used in this manuscript:

\noindent
\setlength{\tabcolsep}{15pt}
\renewcommand{\arraystretch}{1.2}
\begin{tabular}{@{}ll}
CPU & Central Processing Unit\\
GPU & Graphic Processing Unit\\
TPU & Tensor Processing Unit\\
CNN & Convolutional Neural Network\\
ML & Machine Learning\\
MLP & Multi Layer Perceptron\\
AI & Artificial Intelligence\\
PDE & Partial Differential Equation\\
NN & Neural Network\\
GPGPU & Graphics Processing Unit \\
SIMD & Single Instruction Multiple Data \\
SIMT & Single Instruction Multiple Thread \\
FLOPS & Floating Point Operation per Seconds \\
XFS & Extents File System \\
RGB & Red Green Blue\\
EFF & Electronics Frontiers Foundation\\
FPGA & Field Programmable Gate Array \\
DPS & Digital Signal Processing \\
TDP & Thermal Design Power \\
PDF & Protein Data Bank \\
\end{tabular}
\vskip 1cm
\Backmatter
\printbibliography

\end{document}